\documentclass[12pt]{article}
 \usepackage{graphicx}
 \usepackage[cp1251]{inputenc}  
 \usepackage[russian]{babel}    

\def\leaderfill{\leaders\hbox to 1em{\hss.\hss}\hfill}  

\begin{document}

{\small\it
 Astronomy Letters, Vol. 30, No. 11, 2004, pp. 785–796.

 Translated from Pis’ma v Astronomicheski Zhurnal, Vol. 30, No. 11, 2004, pp. 861–873.}

 \bigskip
 \bigskip
 \centerline {\bf KINEMATIC PECULIARITIES OF GOULD BELT STARS}
 \medskip
 \centerline {\bf V.~V.~Bobylev}
 \medskip
 \centerline {\small\it Pulkovo Astronomical Observatory, Russian Academy of Sciences,}
 \centerline {\small\it Pulkovskoe shosse 65, St. Petersburg, 196140 Russia}
 \centerline {\small\it E-mail: vbobylev@gao.spb.ru}
 \medskip
 \medskip
 \centerline {\small Received November 12, 2004}
 \bigskip

{\bf Abstract} We analyzed the space velocities of Gould Belt
stars younger than 125 Myr located at heliocentric distances
 $<650$ pc. We determined the rotation and expansion parameters of the
Gould Belt by assuming the existence of a single kinematic center
whose direction was found to be the following:
 $l_\circ=128^\circ$ and $R_\circ=150$ pc.
The linear velocities reach their maximum at a distance of
 $\approx300$ pc from the center and are
 $-6$ km s$^{-1}$ for the rotation
(whose direction coincides with the Galactic rotation) and
 $+4$ km s$^{-1}$ for the expansion. The stellar rotation model used here
is shown to give a more faithful description of the observed
velocity field than the linear model based on the Oort constants
$A_G$ and $B_G$. We present evidence that the young clusters
$\beta$ Pic, Tuc/HorA, and TWA belong to the Gould Belt structure.


\bigskip
\bigskip
\bigskip
\leftline {\hskip6mm INTRODUCTION}
\bigskip

The star-and-gas complexes associated with star formation
processes in our Galaxy and in other galaxies have a hierarchical
structure (Efremov 1998; Efremov and Elmegreen 1998). Giant
star-and-gas (GSG) complexes trace the spiral pattern of the
Galaxy (Efremov 1998). The GSG complexes have masses of
 $\approx1\times10^6 M_\odot$,
sizes as large as 1000 pc, and lifetimes
 $\tau < 10^8$ yr. Less massive
structures, such as OB associations, open star clusters, and giant
molecular clouds, are part of a GSG complex. The Sun lies within a
GSG complex that is known as the Gould Belt (with a radius of
 $\approx500$ pc and a lifetime of
 $\tau\approx60\times10^6$ yr). In turn, the
Gould Belt is part of an older
 ($\tau\approx5\times10^8$ yr) and more massive
 ($\approx2\times10^7 M_\odot$) structure about 1000 pc in size that is known as the
Local (Orion) Arm or the Local system of stars perceived as a
gravitationally bound, long-lived system. Its kinematics suggests
that it is associated (Olano 2001) with the Sirius star
supercluster (Eggen 1984, 1992). Since the mean residual velocity
of the Sirius supercluster stars relative to the local standard of
rest (LSR) is low, the center of mass of the Local system of stars
has made several turns around the Galactic center in a nearly
circular orbit. Over its lifetime, the Gould Belt has probably
experienced a single impact from a spiral density wave.

An important indicator of the stability of such a system as the
Gould Belt is the existence of proper rotation. Such rotation is
reliably found from observations of Gould Belt stars (Lindblad
2000; Bobylev 2004). Over its lifetime, the Gould Belt has been
affected by the shock wave generated by a spiral density wave at
least once. An analysis of the motions of the OB associations that
fill the interarm space ($r\leq3$ kpc) reveals a complex periodic
structure of their residual velocities attributable to the
influence of spiral density waves (Mel’nik et al. 2001;
Zabolotskikh et al. 2002; Mel’nik 2003). One of these features is
the prominent structure in the radial residual velocity
distribution in Galactocentric distance in the Gould Belt region
(see Fig. 3a in Mel’nik et al. 2001) that we associate with the
well-known positive K-effect or with the expansion of young stars.
In this paper, based on a sample of individual stars that are
members of OB associations, we study this effect in detail.

The goal of this work is to study the motion of the Gould Belt
using data on stars with reliable age estimates. For this purpose,
we use young and nearby OB associations with age estimates
obtained by different authors on the basis of currently available
observations and methods. The rotation and expansion parameters of
the Gould Belt are determined by assuming the existence of a
single kinematic center. To this end, we use Bottlinger’s formulas
in a form that allows the direction of the single kinematic center
to be analytically determined simultaneously with the rotation and
expansion–contraction parameters.

 \bigskip
 \leftline {\hskip6mm METHODS OF ANALYSIS}
 \leftline {\hskip6mm \it Bottlinger’s Formulas}
 \bigskip

In this paper, we use a rectangular Galactic coordinate system
with the axes directed away from the observer toward the Galactic
center
 ($l= 0^\circ, b=0^\circ$, the $X$ axis), along the Galactic rotation
 ($l=90^\circ, b=0^\circ$, the $Y$ axis), and toward the North Galactic Pole
 ($b=90^\circ$, the $Z$ axis). We derived the basic equations from
Bottlinger’s standard formulas (Ogorodnikov 1965). By assuming the
existence of a single kinematic center of rotation and expansion–
contraction ($l_\circ$, $R_\circ$), we transformed the formulas to
$$
\displaylines{\hfill
  V_r= u_G\cos b\cos l+v_G\cos b\sin l+w_G\sin b-\hfill\llap{(1)}
 \cr\hfill
    -D_1 (R-R_\circ)   \sin (l-l_\circ)\cos b-
\hfill\cr\hfill
    -D_2 (R-R_\circ)   \cos (l-l_\circ)\cos b-
\hfill\cr\hfill
    -F_1 (R-R_\circ)^2 \sin (l-l_\circ)\cos b-
\hfill\cr\hfill
    -F_2 (R-R_\circ)^2 \cos (l-l_\circ)\cos b+
\hfill\cr\hfill
  +  k_\circ r \cos^2 b
  + k'_\circ r (R-R_\circ)\cos^2 b+
\hfill\cr\hfill
  + 0.5 k''_\circ r (R-R_\circ)^2 \cos^2 b,
\hfill\cr
\hfill 4.74 r \mu_l\cos b= -u_G\sin l+v_G\cos l+
  \hfill\cr\hfill
  +\omega_\circ r\cos b+ \omega'_\circ r(R-R_\circ)\cos b+
\hfill\cr\hfill
 +0.5\omega''_\circ r\cos b (R-R_\circ)^2-
\hfill\cr\hfill
  - D_1 (R-R_\circ)   \cos (l-l_\circ)
  + D_2 (R-R_\circ)   \sin (l-l_\circ)-
\hfill\cr\hfill
  - F_1 (R-R_\circ)^2 \cos (l-l_\circ)
  + F_2 (R-R_\circ)^2 \sin (l-l_\circ),
\hfill\cr
\hfill 4.74 r \mu_b=-u_G\cos l\sin b- v_G\sin l \sin b +w_G\cos b+
\hfill\cr\hfill
   + D_1 (R-R_\circ)   \sin (l-l_\circ)\sin b+
\hfill\cr\hfill
   + D_2 (R-R_\circ)   \cos (l-l_\circ)\sin b+
\hfill\cr\hfill
   + F_1 (R-R_\circ)^2 \sin (l-l_\circ)\sin b+
\hfill\cr\hfill
   + F_2 (R-R_\circ)^2 \cos (l-l_\circ)\sin b-
\hfill\cr\hfill
  -  k_\circ r \cos b\sin b
  - k'_\circ r (R-R_\circ)\cos b\sin b-
\hfill\cr\hfill
  - 0.5 k''_\circ r (R-R_\circ)^2 \cos b \sin b.
\hfill}
$$
Here, the coefficient 4.74 is the quotient of the number of
kilometers in an astronomical unit by the number of seconds in a
tropical year, $r = 1/\pi$ is the heliocentric distance of the
star, $R_\circ$ is the distance from the Sun to the kinematic
center, and $R$ is the distance from the star to the kinematic
center. Since we corrected the observed motions for the peculiar
solar motion with respect to the LSR (Dehnen and Binney 1998), the
quantities $u_G$, $v_G$, and $w_G$ are (in contrast to the
standard approach, we reversed the signs) the velocity components
of the centroid of the stars under consideration with respect to
the LSR; the components of the stellar proper motion
 $\mu_l \cos b$ and
 $\mu_b$ are in mas yr$^{-1}$
(milliarcseconds per year), the radial velocity $V_r$ is in km
s$^{-1}$, the parallax $\pi$ is in mas (milliarcseconds), and the
distances $R$, $R_\circ$, and $r$ are in kpc. The quantity
$\omega_\circ$ is the angular velocity of the stellar system at
distance $R_\circ$; $k_\circ$ is the radial expansion (or
contraction) rate of the stellar system at distance
 $R_\circ$; $\omega'_\circ$, $\omega''_\circ$,
 and
 $k'_\circ$, $k''_\circ$ are the corresponding derivatives with respect to
the heliocentric distance taken at distance $R_\circ$; and
$l_\circ$ is the direction of the kinematic center. $R$ can be
calculated using the expression
$$
\displaylines{\hfill
 R^2=(r\cos b)^2-2R_\circ r\cos b\cos (l-l_\circ)+R^2_\circ.\hfill\llap(2) }
$$
In Eqs. (1), we use the unknowns
$$
\displaylines{\hfill
  D_1 =  D\cos l_\theta, ~~~
  D_2 = -D\sin l_\theta,\hfill\llap(3)\cr\hfill
  F_1 =  F\cos l_\theta, ~~~
  F_2 = -F\sin l_\theta,\hfill
}
$$
where
 $D=\sqrt{D^2_1+D^2_2}=  \omega' R_\circ$ и
 $F=\sqrt{F^2_1+F^2_2}= 0.5 \omega'' R_\circ$.
Based on relations (3), we determine the correction
 $l_\theta$ to the assumed value of $l_\circ$ twice:
$$
\displaylines{\hfill
  \tan (l_\theta)_D =  -D_2/D_1,\hfill\llap(4)\cr\hfill
  \tan (l_\theta)_F =  -F_2/F_1,\hfill\cr
}
$$
in this case, the new direction is $l_\circ+l_\theta$. On the
other hand, we may introduce similar quantities, $G= k' R_\circ$
and $H= 0.5 k'' R_\circ$, with the unknowns
 $G_1$, $G_2$, $H_1$, and $H_2$
for the expansion to obtain
$$
\displaylines{\hfill
  G_1 =  G\cos l_R= -G\sin l_\theta,\hfill\cr\hfill
  G_2 = -G\sin l_R=  G\cos l_\theta,\hfill\llap(5)\cr\hfill
  H_1 =  H\cos l_R= -H\sin l_\theta,\hfill\cr\hfill
  H_2 = -H\sin l_R=  H\cos l_\theta.\hfill
}
$$
The orthogonality of the expansion and rotation effects
 ($l_\theta=l_R+90^\circ$)
 is reflected in relations (5); as a result, the unknowns in
Eqs. (1) cannot be separated. Therefore, to determine the
direction of the kinematic center, it will suffice to use one of
the sets of unknowns, for example, $D_1$, $D_2$, $F_1$, and $F_2$,
in Eqs. (1). Based on the described approach, we can independently
estimate the distance from the Sun to the kinematic center. This
estimate can be obtained from the formulas that follow from
relations (3)–(5):
$$
\displaylines{\hfill
  D =     R_\circ|\omega'_\circ-k'_\circ|,\hfill\llap(6)\cr\hfill
  F = 0.5 R_\circ|\omega''_\circ-k''_\circ|.\hfill
}
$$
Equations (1) contain thirteen sought unknowns:
 $u_{\odot}$, $v_{\odot}$, $w_{\odot}$, $\omega_\circ$, $\omega'_\circ$,
 $\omega''_\circ$, $k_\circ$, $k'_\circ$, $k''_\circ$, $D_1$,
 $D_2$, $F_1$, $F_2$,
 which can be determined by the least-squares method.
 If the direction of the kinematic center is known, then
the equations have the original form
$$
\displaylines{\hfill
  V_r= u_G\cos b\cos l+\hfill\llap{(7)}
\cr\hfill
      +v_G\cos b\sin l+w_G\sin b
      -R_\circ (R-R_\circ)\times
\hfill\cr\hfill
  \times \sin (l-l_\circ) \cos b \omega'_\circ
  -0.5 R_\circ (R-R_\circ)^2 \times
\hfill\cr\hfill
  \times\sin (l-l_\circ)\cos b \omega''_\circ
  +\cos^2 b k_\circ r+
\hfill\cr\hfill
  +(R-R_\circ)(r\cos b-R_\circ\cos (l-l_\circ))\times
\hfill\cr\hfill
  \times\cos b k'_\circ+0.5(R-R_\circ)^2(r\cos b-
\hfill\cr\hfill -R_\circ\cos (l-l_\circ))\cos b k''_\circ,
\hfill\cr \cr
\hfill 4.74 r \mu_l\cos b= -u_G\sin l+ \hfill\cr\hfill
  +v_G\cos l-(R-R_\circ)(R_\circ\cos (l-l_\circ)-
\hfill\cr\hfill
 -r\cos b)\omega'_\circ-0.5 (R-R_\circ)^2(R_\circ\cos (l-l_\circ)-
\hfill\cr\hfill
 -r\cos b)\omega''_\circ+r\cos b \omega_\circ+R_\circ(R-R_\circ)\times
\hfill\cr\hfill
 \times\sin (l-l_\circ)) k'_\circ+0.5R_\circ(R-R_\circ)^2\times
\hfill\cr\hfill \times\sin (l-l_\circ)) k''_\circ,\hfill \cr \cr
\hfill 4.74 r \mu_b=-u_G\cos l\sin b- \hfill\cr\hfill
 -v_G\sin l \sin b +w_G\cos b+
 R_\circ(R-R_\circ)\times
\hfill\cr\hfill
  \times\sin (l-l_\circ)\sin b \omega'_\circ
  +0.5 R_\circ (R-R_\circ)^2\times
\hfill\cr\hfill
  \times \sin (l-l_\circ)\sin b \omega''_\circ
  -\cos b\sin b k_\circ r-
\hfill\cr\hfill
  -(R-R_\circ)(r\cos b-R_\circ\cos (l-l_\circ))\times
\hfill\cr\hfill \times\sin b k'_\circ-0.5(R-R_\circ)^2(r\cos b-
\hfill\cr\hfill -R_\circ\cos(l-l_\circ))\sin b k''_\circ. \hfill}
$$
These equations contain nine sought unknowns:
 $u_{\odot}$, $v_{\odot}$, $w_{\odot}$, $\omega_\circ$, $\omega'_\circ$,
 $\omega''_\circ$, $k_\circ$, $k'_\circ$, $k''_\circ$.
A peculiarity of this method is that it requires the existence of
derivatives only with respect to the distance. A sample of stars
uniformly distributed even in one Galactic quadrant can satisfy
this requirement.

 \bigskip
 \leftline {\hskip6mm \it The Statistical Method}
 \bigskip

We use the standard statistical method (Tr\"umpler and Weaver
1953; Parenago 1954; Ogorodnikov 1965) that consists in
determining and analyzing the symmetric moment tensor or the
stellar residual velocity dispersion tensor. When both the radial
velocities and proper motions of stars are used to determine the
six unknown components of the dispersion tensor, we have six
equations that can be written for each star. The semiaxes of the
residual velocity (Schwarzschild) ellipsoid that we denote by
 $\sigma_{1,2,3}$ can be determined by analyzing the eigenvalues of the
dispersion tensor. We denote the directions of the principal axes
of this ellipsoid by $l_{1,2,3}$ and $b_{1,2,3}$. A peculiarity of
the approach used here is that the stellar velocities corrected
for the peculiar solar motion with respect to the LSR and for the
general Galactic rotation are used as the residual velocities.

 \bigskip
 \leftline {\hskip6mm \it The Ogorodnikov–Milne Model}
 \bigskip

In the linear Ogorodnikov–Milne model, we use the same notation
that was introduced by Clube (1972, 1973) and used by du Mont
(1977, 1978). We modify the linear model developed to describe the
general Galactic rotation to describe the peculiarities of the
velocity field of nearby stars. To within terms of the first order
of smallness $r/R_\circ\ll1$, the observed velocity ${\bf V}(r)$
of a star with a heliocentric radius vector r is described by the
vector equation
$$
\displaylines{ \hfill
 {\bf V}(r)-{\bf V}_{GR}-{\bf V_\odot}_{LSR}={\bf V}_G+M{\bf r}+{\bf V'}.\hfill\llap(8)\cr}
$$
Here,
 ${\bf V}_{GR}$ is the systematic velocity of the star attributable
to the general Galactic rotation,
 ${\bf V}_{LSR}$ is the peculiar solar
motion with respect to the LSR,
 ${\bf V}_G (u_G, v_G,w_G)$ is the velocity of
the centroid of the stars under consideration relative to the LSR,
 ${\bf V}'$ is the residual stellar velocity (the residual stellar
velocities are assumed to be distributed randomly), and
 $M$ is the
displacement matrix that describes systematic motions similar to
the proper rotation and expansion–contraction effects. The
components of the matrix M are the partial derivatives of
 ${\bf u} (u_1,u_2,u_3)$ with respect to
 ${\bf r} (r_1,r_2,r_3)$:
$$
\displaylines{\hfill M_{pq}={\left(\frac{\partial u_p} {\partial
r_q}\right)}_\circ, \quad (p,q=1,2,3). \hfill\llap(9)\cr }
$$
The matrix $M$ can be separated into symmetric, $M^+$, and
antisymmetric, $M^-$, parts. Following Ogorodnikov (1965), we call
them the local deformation and local rotation tensors,
respectively:
$$
\displaylines{\hfill M_{\scriptstyle
pq}^{\scriptscriptstyle+}={1\over 2}\left( \frac{\partial
u_{p}}{\partial r_{q}}+ \frac{\partial u_{q}}{\partial
r_{p}}\right)_\circ,
  \hfill\llap(10)\cr\hfill
M_{\scriptstyle pq}^{\scriptscriptstyle-}={1\over 2}\left(
\frac{\partial u_{p}}{\partial r_{q}}- \frac{\partial
u_{q}}{\partial r_{p}}\right)_\circ, \hfill
\cr\hfill(p,q=1,2,3).\hfill }
$$
The basic equations are
$$
\displaylines{\hfill
 V_r= u_G \cos b\cos l+v_G\cos b\sin l+w_G\sin b+
\hfill\llap(11)\cr\hfill
 +r (\cos^2 b\cos^2 l M_{11}
 +\cos^2 b\cos l \sin l M_{12}+
\hfill\cr\hfill
 +\cos b\sin b \cos l  M_{13}
 +\cos^2 b\sin l\cos l M_{21}+
\hfill\cr\hfill
 +\cos^2 b\sin^2 l   M_{22}
 +\cos b\sin b\sin l M_{23}+
\hfill\cr\hfill
 +\sin b\cos b\cos l M_{31}
 +\cos b\sin b\sin l M_{32}+
\hfill\cr\hfill
 +\sin^2 b  M_{33}),
\hfill\cr \cr
 \hfill
 4.74 r \mu_l\cos b= -u_G\sin l+v_G\cos l+ ~~~
\hfill \cr\hfill
 +r (-\cos b\cos l\sin l  M_{11}
 -\cos b\sin^2 l  M_{12}-
\hfill\cr\hfill
 -\sin b \sin l  M_{13}
 +\cos b\cos^2 l M_{21}+
\hfill\cr\hfill
 +\cos b\sin l\cos l  M_{22}+
\hfill\cr\hfill
 +\sin b\cos l  M_{23}),
\hfill\cr \cr
 \hfill
4.74 r \mu_b=-u_G\cos l\sin b-v_G\sin l\sin b+w_G\cos b+
\hfill\cr\hfill
 +r (-\sin b\cos b\cos^2 l M_{11}-
\hfill\cr\hfill
 -\sin b\cos b\sin l \cos l M_{12}-
\hfill\cr\hfill
 -\sin^2 b \cos l  M_{13}
 -\sin b\cos b\sin l\cos l M_{21}-
\hfill\cr\hfill
 -\sin b\cos b\sin^2 l  M_{22}
 -\sin^2 b\sin l  M_{23}+
\hfill\cr\hfill
 +\cos^2 b\cos l M_{31}
 +\cos^2 b\sin l M_{32}+
\hfill\cr\hfill
 +\sin b\cos b  M_{33}),
\hfill }
$$
where the stellar velocity components corrected for the general
Galactic rotation and for the peculiar solar motion with respect
to the LSR appear on the left-hand sides. Equations (11) contain
twelve sought unknowns: the three velocity components
 $V_G(u_G,v_G,w_G)$ and the nine
components
 $M_{pg}$ that can be determined by the least-squares
method. The deformation and rotation tensor components can be
calculated using the values of Mpq derived from relations (10).
This method requires the existence of all derivatives with respect
to the coordinates. A sample of stars that densely fill each
coordinate axis is needed to meet this requirement.

\bigskip
 \leftline {\hskip6mm OBSERVATIONAL MATERIAL}
 \leftline {\hskip6mm \it Data on Stars}
\bigskip

We compiled our working list of stars located no farther than
 $\approx650$ pc
 from the Sun using the following sources: we took the Hipparcos
(ESA 1997) star numbers for the Col 121, Per OB2, Vel OB2, Tr 10,
LCC, UCL, US, Cep OB2, Lac OB1, Cep OB6, $\beta$ Per (Per OB3),
and Cas Tau associations from de Zeeuw et al. (1999); for the Ori
OB1 a–d associations from Brown et al. (1994); for the IC 2391, IC
2602, NGC 2232, NGC 2451, NGC 2516, NGC 2547, and Pleiades
clusters from the list by Robichon et al. (1999); the list of
members of the TW Hydrae (TWA),  $\beta$ Pic, and Tucana/HorA
associations from Song et al. (2003); and for the a Car cluster
from Platais et al. (1998). The data on 24 stars having X-ray
emissions and located no farther than 50 pc from the Sun were
taken from Makarov (2003), where they are denoted as an XY sample.
The star HIP 30030 in Makarov’s list is a member of the TWA
cluster, belongs to the Tucana/HorA group, and is designated TWA42
(Song et al. 2003). The data on 34 stars located no farther than
100 pc from the Sun were taken from Wichmann et al. (2000, 2003).
Four stars were not included in other lists; we placed them in
group 2. Table 1 gives the ages of the selected clusters. We took
the equatorial coordinates, parallaxes, and proper motions from
the Hipparcos catalog and the radial velocities from the catalog
by Barbier-Brossat and Figon (2000). We use only single stars (the
astrometric orbital binaries marked by the symbol O were rejected)
for which the parallaxes, radial velocities, and proper motions
are available. Based on the stellar ages (Table 1), we divided the
stars into three groups: (1) the youngest stars with ages $<10$
Myr (group 1); (2) the stars with middle ages of
 $10-60$ Myr (group 2); and (3) the old stars with ages of
 $60-125$ Myr (group 3).

 \bigskip
 \leftline {\hskip6mm \it Allowance for the Galactic Rotation}
 \bigskip

We took into account the general Galactic rotation using the
parameters found previously (Bobylev 2004):
 $\omega_\circ  = -28.0\pm0.6$ km s$^{-1}$ kpc$^{-1}$,
 $\omega'_\circ = +4.17\pm0.14$ km s$^{-1}$ kpc$^{-2}$, and
 $\omega''_\circ= -0.81\pm0.12$ km s$^{-1}$ kpc$^{-3}$. The Galactocentric distance of
the Sun was assumed to be $R_\circ=7.1$ kpc, which corresponds to
the short distance scale (Dambis et al. 2001). The radial
velocities and proper motions of the stars were corrected for
peculiar solar motion with respect to the LSR using the values
obtained by Dehnen and Binney (1998):
 $(u_\odot, v_\odot, w_\odot) = (10.0, 5.3, 7.2)$ km s$^{-1}$.

 \bigskip
 \leftline {\hskip6mm RESULTS}
 \bigskip

A preliminary analysis of the stellar space velocities showed the
necessity of setting a limit on the total stellar velocity,
 $\sqrt{U^2+V^2+W^2}<30$ km s$^{-1}$,
 which we use below. The residual velocities
$U$, $V$, and $W$ were calculated from standard formulas
(Kulikovskii 1985). We used stars with parallaxes $\pi>1.5$ mas
($r<667$ pc). The number of stars used is given in the last column
of Table 1. Thus, each of groups 1, 2, and 3 always contains a
fixed number of stars, 114, 342, and 150, respectively. For our
sample of stars, the mean error in the parallax is $15-17\%$. The
mean errors in the proper motion components of the stars
 $4.74r\mu_l\cos b$ and $4.74r\mu_b$,
including the
errors in the parallaxes, are
 $\approx1$ km s$^{-1}$. For
 $\approx40\%$ of
the stars, there are no data on the errors in the radial
velocities (Barbier-Brossat and Figon 2000); for stars with
available information about the errors, the mean error is
 $\approx3.5$ km s$^{-1}$.

 The group 1 stars are distributed exclusively in
Galactic quadrant III. The group 2 stars fill the solar
neighborhood within $r\approx650$ pc of the Sun most uniformly.
The group 3 stars densely fill a compact zone within $r\approx50$
pc of the Sun in Galactic quadrant II. We also formed a combined
group of young ($<60$ Myr) stars composed of the group 1 and group
2 stars. For the latter combined group, the application of all
kinematic models proved to be possible.

As the first approximation for the direction of the kinematic
center, we take $l_\circ=160^\circ$ and $R_\circ=150$ pc (Bobylev
2004). Solving the system of equations (1) for the combined group
of young stars (group~1+group~2) yielded the following kinematic
parameters of the linear motion of the stars:
 $(u_G, v_G,w_G)=(0.0\pm0.7, -12.3\pm0.7, 1.3\pm0.3)$ km s$^{-1}$.
 Consequently, the stars under
consideration move with respect to the LSR at a velocity of
 $V_G = 12.3\pm0.7$ km s$^{-1}$ in the direction
 $L_G = 270\pm3^\circ$ and
 $B_G = 6\pm1^\circ$.
Further,
$$
\displaylines{\hfill
 \omega_\circ  = -32.4\pm  3.8~\hbox{km s$^{-1}$ kpc$^{-1}$},    \hfill\llap(12)\cr\hfill
 \omega'_\circ = +93.2\pm 28.3~\hbox{km s$^{-1}$ kpc$^{-2}$},\hfill\cr\hfill
\omega''_\circ=-170.5\pm107.8 ~\hbox{km s$^{-1}$ kpc$^{-3}$},
\hfill\cr\hfill
      k_\circ = +27.9\pm  3.8 ~\hbox{km s$^{-1}$ kpc$^{-1}$},    \hfill\cr\hfill
     k'_\circ =-122.3\pm 28.3 ~\hbox{km s$^{-1}$ kpc$^{-2}$},\hfill\cr\hfill
    k''_\circ =+233.7\pm107.8 ~\hbox{km s$^{-1}$ kpc$^{-3}$},\hfill\cr\hfill
       l_\circ=   128^\circ,\hfill\cr\hfill
 (l_\theta)_D =     2^\circ,\hfill\cr\hfill
 (l_\theta)_F =   181^\circ,\hfill\cr\hfill
  (R_\circ)_D =   180~\hbox{pc},\hfill\cr\hfill
  (R_\circ)_F =   410~\hbox{pc}.\hfill
}
$$
 %
 %
 %


The results of the solution of (12) are presented in Fig.~1, where
line 1 indicates the derived expansion curve with the estimated
parameters $k_\circ$, $k'_\circ$, and $k''_\circ$; line 2
indicates the rotation curve with the estimated parameters
 $\omega_\circ$, $\omega'_\circ$, and $\omega''_\circ$.
As we see from the solution of (12), the value of $R_\circ=180$ pc
calculated using the first derivatives is in satisfactory
agreement with the $R_\circ=150$ pc found previously (Bobylev
2004). The expansion curve constructed by solving (12) has a
velocity maximum of $V=4.4$ km s$^{-1}$ at $R=200$ pc from the
center. Based on these data, we determined the time elapsed from
the onset of expansion, $\tau=44$ Myr, and it is in good agreement
with the ages of the stars under consideration.

Applying the Ogorodnikov–Milne model (Eqs. (11)) to the combined
group of young stars (group~1+group~2), we found the following
components of the tensors
 $M$, $M^+$, and $M^-$ (km s$^{-1}$ kpc$^{-1}$):
$$
\displaylines{\hfill M=\pmatrix
 { 13.0_{(3.3)}&  8.1_{(1.4)}& -4.4_{(8.6)}\cr
  -21.5_{(3.3)}&  2.9_{(1.4)}& -7.0_{(8.6)}\cr
    4.6_{(3.3)}&  4.3_{(1.4)}&-18.8_{(8.6)}\cr},\hfill\llap(13)\cr\hfill
M^{\scriptscriptstyle+}=\pmatrix
 { 13.0_{(3.3)}& -6.7_{(1.8)}&  0.1_{(4.6)}\cr
   -6.7_{(1.8)}&  2.9_{(1.4)}& -1.4_{(4.3)}\cr
    0.1_{(4.6)}& -1.4_{(4.3)}&-18.8_{(8.6)}\cr},\hfill\cr\hfill
M^{\scriptscriptstyle-}=\pmatrix
  {           0& 14.8_{(1.8)}& -7.1_{(5.0)}\cr
  -14.8_{(1.8)}&            0& -4.5_{(4.6)}\cr
    4.5_{(4.6)}&  5.7_{(4.3)}&            0\cr}.\hfill
}
$$
The Oort parameters are:
 $A_G=M^+_{21}=-6.7\pm1.8$ km s$^{-1}$ kpc$^{-1}$,
 $B_G=M^-_{21}=-14.8\pm1.8$ km s$^{-1}$ kpc$^{-1}$,
 $C_G=0.5(M^+_{11}-M^+_{22})=5.1\pm1.8$ km s$^{-1}$ kpc$^{-1}$,
 $K_G=0.5(M^+_{11}+M^+_{22})=8.0\pm1.8$ km s$^{-1}$ kpc$^{-1}$.
 In Eqs. (11), the $X$ axis is directed
toward the Galactic center ($l_\circ=0^\circ$). The deformation
tensor in the principal axes $M^+_G$ is (km s$^{-1}$ kpc$^{-1}$):
$$
\displaylines{ M^{\scriptscriptstyle+}_G=
 \pmatrix
 { 16.4&  0  &  0   \cr
      0& -0.4&  0   \cr
      0&  0  &-18.9 \cr};
}
$$
the directions of the principal axes are:
$$
\displaylines{\hfill
 L_1=153.5\pm0.2^\circ,\quad B_1=-1.1\pm0.0^\circ, \hfill\llap(14)\cr\hfill
 L_2=  243\pm 10^\circ,\quad B_2=   4\pm  8^\circ, \hfill\cr\hfill
 L_3=   81\pm 10^\circ,\quad B_3=   86\pm 11^\circ.\hfill}
$$
The vertex deviation in the $XY$ plane that we determined from the
formula $\tan 2l_{XY}=-C/A$ is $l_{XY}=19\pm6^\circ$ and indicates
one of the directions of the rotation center; the other direction
is $l=109^\circ$ and is close to the direction of the kinematic
center $l_\circ=128^\circ$ that we found using Eqs. (1).

An analysis of the deformation tensor (19), (20) leads us to the
important conclusion (which cannot be drawn from model (1)) that
expansion takes place exclusively in the Galactic $XY$ plane.

The results of applying the statistical method are given in
Table~2. As we see from this table, the direction of the rotation
center for the combined group of stars (Group~1+Group~2) is
determined accurately, $l_1=132.7\pm0.4^\circ$. The first axis of
the residual velocity ellipsoid (the vertex deviation) changes its
direction with the age of the stars under consideration. This is
consistent with the dynamic model of
 %
 %
 %
 %
Olano (2001). As Olano showed, the direction of the vertex
rotation depends on the critical stellar density in the solar
neighborhood $\rho/\rho^*$, where $\rho$ is given in
 $M_\odot$~pc$^{-3}$.
Following Olano’s estimates ($\rho/\rho^*=1.5$), we assume that
the change occurs counterclockwise in the coordinate system under
consideration. Thus, the vertex deviation changes from
 $l_1=36^\circ$ for the sample of old stars (Group 3) to
 $l_1=131^\circ$ for the sample of middle-aged stars (Group 2) and to
 $l_1=172^\circ$ for the sample of the youngest stars (Group 1). Associating
this direction with the direction of the kinematic center, we can
see that the center does not remain constant. Therefore, its
independent determination is required for each sample of stars.
There is good agreement of the orientation of the residual
velocity tensor for the Group 1 and Group 2 stars that we found
(Table 2) with the geometric characteristics of the Gould Belt
(Torra et al. 2000). For example, for the group-2 stars, the
directions
 $l_2= 224\pm18^\circ$ and
 $b_2= -22\pm 5^\circ$ almost match the
characteristics of the Gould Belt
 $\Omega_G=275-295^\circ$ and $i_G=16-20^\circ$
(here, $\Omega_G$ is the direction of the line of nodes of the
Gould Belt and $i_G$ is the inclination of the disk to the
Galactic plane) found by Torra et al. (2000).

For the combined group of stars (Group~1+Group~2), we obtained the
solutions of Eqs. (7) in which we used
 $l_\circ=128^\circ$ and $R_\circ=150$ pc. Equations (7) were solved
for two cases, with nine and seven unknowns in the first and
second cases, by assuming that the second derivatives
 $\omega''_\circ$ and
 $k''_\circ$ are equal to zero. The results are presented in Table 3 and
Fig. 2. In Fig. 2, curves 1 and 2 were constructed using the data
in the upper part of Table. 3; curves 5 and 6 were constructed
using the data in the lower part of Table 3 (seven unknowns).
Curve 4 in Fig. 2 represents the results of applying the
Oort–Lindblad model, which for
 $l_\circ=128^\circ$ are the following:
 $A_G=  6.5\pm1.8$ km s$^{-1}$ kpc$^{-1}$ and
 $B_G=-14.8\pm1.8$ km s$^{-1}$ kpc$^{-1}$ (these
values were obtained from the equations of the Ogorodnikov– Milne
model, but the curve is essentially the Oort–Lindblad
approximation). The time elapsed from the onset of expansion for
curves 2 and 6 is about
 $\tau=120$ Myr, which significantly exceeds
the estimate obtained by solving (12).

The results of solving Eqs. (7) for group 3 are given in the lower
part of Table 3. The parameters derived for group 3 differ
significantly from those of young stars: the rotation and the
expansion have opposite signs. We associate the kinematic
peculiarities of the Local system of stars that were found by
Tsvetkov (1995a, 1995b, 1997) by studying type-A and -F stars with
these stars. It is the influence of the Group 3 stars that led to
the paradox that younger OB stars showed a slower rotation than
older stars (Bobylev 2004).

Figures 3 and 4 show the residual velocities $U$, $V$, and $W$ of
middle-aged Gould Belt stars in projection onto the Galactic $XY$
and $XZ$ planes. These velocities are residual in every sense,
because they were corrected from both the peculiar solar motion
and the total velocity $V_G$ that we found. Thus, we relate the
 %
 %
 %
 %
velocities to a coordinate system whose center is the center of
mass of the Gould Belt that moves with respect to the LSR.

Figure~5 shows the residual velocities of stars in the young
$\beta$ Pic and Tuc/HorA clusters together with TWA cluster stars.

\bigskip
\leftline {\hskip6mm DISCUSSION}
\bigskip

The kinematic parameters that we derived using the linear
Ogorodnikov–Milne model are in excellent agreement with the
kinematic model of the Gould Belt suggested by Lindblad (2000),
which is based on an analysis of the results by Comeron (1999) and
Torra et al. (1997) for the age interval of Gould Belt stars
$20-40$ Myr. Lindblad’s model assumes the proper rotation of the
Gold Belt stars with an angular velocity of
 $\omega_G = B-A_\omega = -24$ km s$^{-1}$ kpc$^{-1}$
 ($A_\omega= 6.4$ km s$^{-1}$ kpc$^{-1}$ and
  $B=-17.4$ km s$^{-1}$ kpc$^{-1}$) and
the expansion with
 $K=11.3$ km s$^{-1}$ kpc$^{-1}$ in the direction of the
center
 $l_\circ=127^\circ$. An analysis of the curves in Fig. 2 indicates
that the linear model faithfully describes the observed velocity
field of the stars under consideration no farther than $200-300$
pc from the kinematic center. Our rotation parameters for the
Gould Belt are consistent with the model by Olano (2001) and with
the evolution model of the youngest moving clusters by Asiain et
al. (1999). The question of whether the $\beta$ Pic, Tuc/HorA, and
TWA clusters belong to the Gould Belt structure is currently being
debated. Makarov and Fabricius (2001) directly associate these
clusters with the Gould Belt; Song et al. (2003) cast doubt on
this association. In our opinion, a comparison of the stellar
residual velocities for all of the clusters ($\beta$ Pic,
Tuc/HorA, and TWA) in Fig. 5 with our rotation and expansion
curves suggests that the cluster stars belong to the Gould Belt
structure. The last row in Table 2 gives our calculated parameters
of the residual velocity ellipsoid for all 33 stars of the $\beta$
Pic, Tuc/HorA, and TWA clusters. An analysis of
 $l_1$, $b_1$, $l_2$, $b_2$, $l_3$, $b_3$
and a comparison of Figs. 4 and 5 indicate that these stars belong
to the disk of the Gould Belt. Ortega et al. (2002) performed
model calculations of the orbits of $\beta$ Pic cluster members in
the Galactic field of attraction for an interval of 11.5 Myr and
used them to determine the birthplace of the cluster. As follows
from an analysis of our rotation curve for the Gould Belt (line 1
in Fig.~2), the Gould Belt has a substantial mass. Therefore, in
our view, to determine the birthplaces of the $\beta$ Pic,
Tuc/HorA, and TWA clusters, the gravitational potential of the
Gould Belt
 ($\approx2\times10^6 M_\odot$) or the entire Local system of stars
 ($\approx2\times10^7 M_\odot$) must be taken into account.

\bigskip
\leftline {\hskip6mm CONCLUSIONS}
\bigskip

We have analyzed the residual velocity field of nearby and young
Gould Belt stars by assuming the existence of a single kinematic
center (the center of proper rotation and expansion). The
direction of the kinematic center of the Gould Belt was found to
be: $l_\circ=128^\circ$ and $R_\circ=150$ pc.

Using the Ogorodnikov–Milne model, we showed that the vectors of
the radial component of the residual velocity field (a positive
 $K$-effect) lie in the Galactic $XY$ plane unrelated to the plane of
symmetry of the Gould Belt disk. This led us to conclude that the
effect in question results from the impact of a spiral density
wave propagating precisely through the Galactic disk on the cloud
of gas out of which the stars under consideration were
subsequently formed. In contrast, the effect of proper rotation is
closely related to the plane of symmetry of the Gould Belt disk.

We have considered an approach that, in our view, allows the
observed residual velocity field of the Gould Belt stars to be
described more faithfully than does the linear (Oort–Lindblad)
model. The linear velocities that we found for a sample of stars
younger than
 $60$ Myr reach their maximum at a distance of $\approx$300 pc from the
kinematic center and are
 $-6$ km s$^{-1}$ for the rotation and $+4$ km s$^{-1}$
for the expansion. Our constructed proper rotation curve for the
sample of young stars in the Gould Belt suggests that its mass is
substantial. This conclusion is in agreement with the spatial
distribution of faint stars with X-ray emission (Guillout et al.
1998). We argue that the young $\beta$ Pic, Tuc/HorA, and TWA
clusters belong to the Gould Belt structure.

\bigskip
\leftline {\hskip6mm ACKNOWLEDGMENTS}
\bigskip

I am grateful to A.~T.~Bajkova for her help and useful
discussions. This work was supported by the Russian Foundation for
Basic Research (project no. 02-02-16570).

\bigskip
\leftline {\hskip6mm REFERENCES}

1. R. Asiain, F. Figueras, and J. Torra, Astron. Astrophys.
 {\bf 350}, 434 (1999).

2. M. Barbier-Brossat and P. Figon, Astron. Astrophys., Suppl.
Ser. {\bf 142}, 217 (2000).

3. V. V. Bobylev, Pis’ma Astron. Zh. {\bf 30}, 185 (2004) [Astron.
Lett. {\bf 30}, 159 (2004)].

4. A. G. A. Brown, E. J. de Geus, and P. T. de Zeeuw, Astron.
Astrophys. {\bf 289}, 101 (1994).

5. F. Carrier, G. Burki, and C. Richard, Astron. Astrophys. {\bf
341}, 469 (1999).

6. S. V. M. Clube, Mon. Not. R. Astron. Soc. {\bf 159}, 289
(1972).

7. S. V. M. Clube, Mon. Not. R. Astron. Soc. {\bf 161}, 445
(1973).

8. F. Comer\'on, Astron. Astrophys. {\bf 351}, 506 (1999).

9. A. K. Dambis, A. M. Mel’nik, and A. S. Rastorguev, Pis’ma
Astron. Zh. {\bf 27}, 68 (2001) [Astron. Lett. {\bf 27}, 58
(2001)].

10. W. Dehnen and J. J. Binney, Mon. Not. R. Astron. Soc. {\bf
298}, 387 (1998).

11. P. T. de Zeeuw, R. Hoogerwerf, J. H. J. de Bruijne, et al.,
Astron. J. {\bf 117}, 354 (1999).

12. B. du Mont, Astron. Astrophys. {\bf 61}, 127 (1977).

13. B. du Mont, Astron. Astrophys. {\bf 66}, 441 (1978).

14. Y. N. Efremov, Astron. Astrophys. Trans. {\bf 15}, 3 (1998).

15. Y. N. Efremov and B. G. Elmegreen, Mon. Not. R. Astron. Soc.
{\bf 299}, 588 (1998).

16. O. J. Eggen, Astron. J. {\bf 89}, 1350 (1984).

17. O. J. Eggen, Astron. J. {\bf 104}, 1493 (1992).

18. A. Gim\'enez and J. V. Clauzen, Astron. Astrophys. {\bf 291},
795 (1994).

19. P. Guillout, M. F. Sterzik, J. H.M.M. Schmitt, et al., Astron.
Astrophys. {\bf 337}, 113 (1998).

20. R. D. Jeffries and A. J. Tolley, Mon. Not. R. Astron. Soc.
{\bf 300}, 331 (1998).

21. P. G. Kulikovskii, {\it Stellar Astronomy} (Nauka, Moscow,
1985) [in Russian].

22. K. L. Luhman, Astrophys. J. {\bf 560}, 287 (2001).

23. P. O. Lindblad, Astron. Astrophys. {\bf 363}, 154 (2000).

24. V. V. Makarov, Astron. J. {\bf 126}, 1996 (2003).

25. V. V. Makarov and C. Fabricius, Astron. Astrophys. {\bf 368},
866 (2001).

26. E. E. Mamajek, M. Meyer, and J. Liebert, Astron. J. {\bf 124},
1670 (2002).

27. A. M. Mel’nik, Pis’ma Astron. Zh. {\bf 29}, 349 (2003)
[Astron. Lett. {\bf 29}, 304 (2003)].

28. A. M. Mel’nik, A. K. Dambis, and A. S. Rastorguev, Pis’ma
Astron. Zh. {\bf 27}, 611 (2001) [Astron. Lett. {\bf 27}, 521
(2001)].

29. D. Montes, J. Lo\'pez-Santiago,M. C. G\'alvez, et al., Mon.
Not. R. Astron. Soc. {\bf 328}, 45 (2001).

30. K. F. Ogorodnikov, {\it Dynamics of Stellar Systems}
(Fizmatgiz, Moscow, 1965) [in Russian].

31. C. A. Olano, Astron. J. {\bf 121}, 295 (2001).

32. V. G. Ortega, de la Reza, E. Jilinski, et al., Astrophys. J.
Lett. {\bf 575}, L75 (2002).

33. P. P. Parenago, {\it A Course on Stellar Astronomy} (Gosizdat,
Moscow, 1954) [in Russian].

34. I. Platais, V. Kozhurina-Platais, and F. van Leeuwen, Astron.
J. {\bf 116}, 2423 (1998).

35. S. Randich, N. Aharpour, R. Pallavicini, et al., Astron.
Astrophys. {\bf 323}, 86 (1997).

36. N. Robichon, F. A. Arenou, J.-C. Mermilliod, et al., Astron.
Astrophys. {\bf 345}, 471 (1999).

37. M. J. Sartori, J. R. D. Leґ pine, andW. S. Dias, Astron.
Astrophys. {\bf 404}, 913 (2003).

38. D. R. Soderblom, B. F. Jones, S. Balachandran, et al., Astron.
J. {\bf 106}, 1059 (1993).

39. I. Song, B. Zuckermann, and M. S. Bessel, Astrophys. J. {\bf
599}, 342 (2003).

40. J. R. Stauffer, L. W. Hartmann, C. F. Prosser, et al.,
Astrophys. J. {\bf 479}, 776 (1997).

41. The Hipparcos and Tycho Catalogues, ESA SP-1200 (1997).

42. J. Torra, D. Fern aґ ndez, and F. Figueras, Astron. Astrophys.
{\bf 359}, 82 (2000).

43. J. Torra, A. E. G\'omez, F. Figueras, et al.,
 {\it HIPPARCOS Venice’97},
 Ed. by B. Battrick (ESA Publ. Div., Noordwijk, 1997), p. 513.

44. R. J. Tr\"umpler and H. F.Weaver, {\it Statistical Astronomy}
(Univ. of California Press, Berkely, 1953).

45. A. S. Tsvetkov, Astron. Astrophys. Trans. {\bf 8}, 145 (1995).

46. A. S. Tsvetkov, Astron. Astrophys. Trans. {\bf 9}, 1 (1995).

47. A. Tsvetkov, {\it JOURNEES-97}, Ed. by J. Vondr\'ak and N.
Capitaine (Observatoire de Paris, Paris, 1997), p. 171.

48. R. Wichmann, J. H. M. M. Schmitt, and S. Hubrig, Astron.
Astrophys. {\bf 399}, 983 (2003).

49. R. Wichmann, G. Torres, C. H. F. Melo, et al., Astron.
Astrophys. {\bf 359}, 181 (2000).

50. M. V. Zabolotskikh, A. S. Rastorguev, and A. K. Dambis, Pis’ma
Astron. Zh. {\bf 28}, 516 (2002) [Astron. Lett. {\bf 28}, 454
(2002)].


\begin{figure}[t]
{\begin{center}
  \includegraphics[width=160mm]{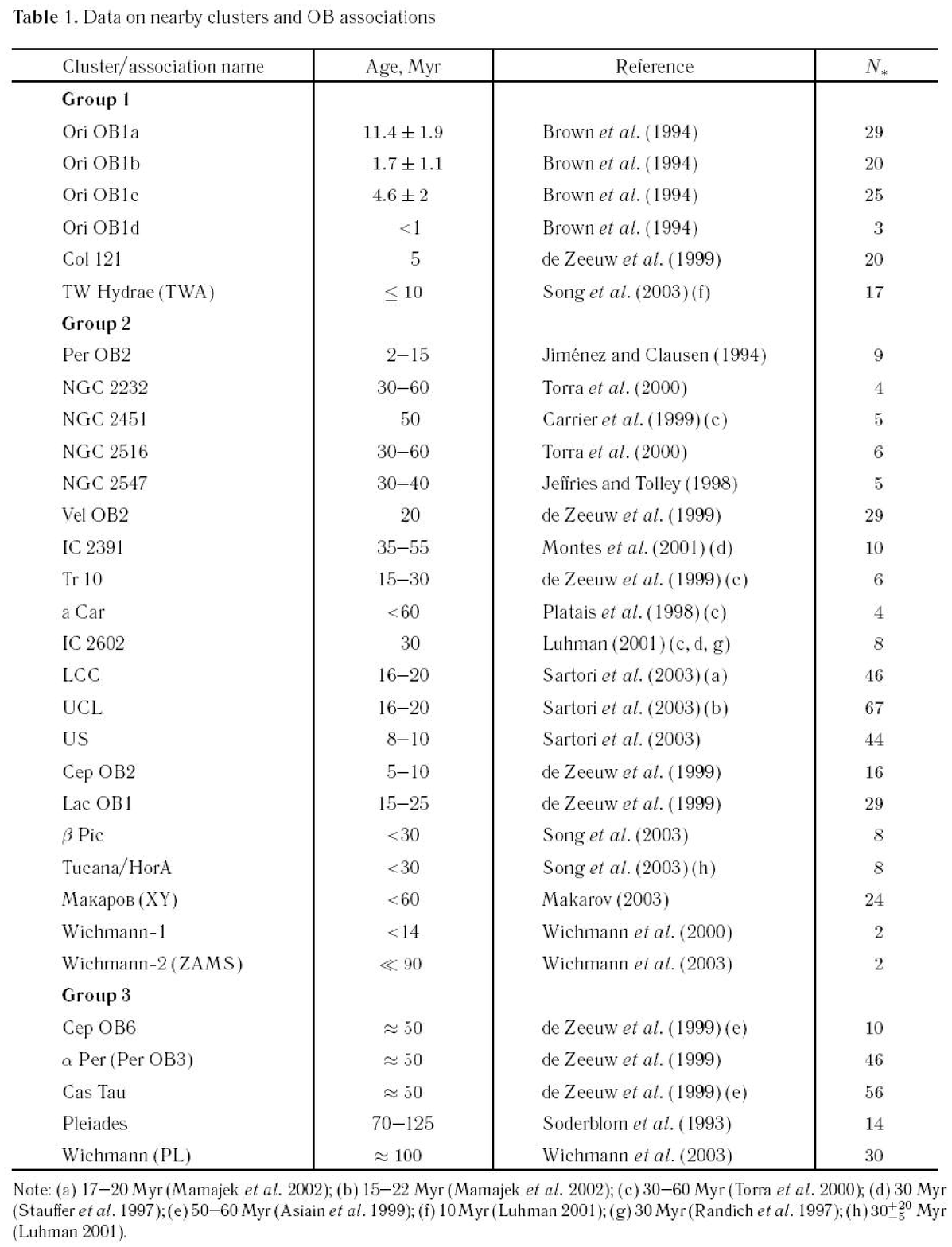}
\end{center}}
\end{figure}

\begin{figure}[t]
{\begin{center}
  \includegraphics[width=160mm]{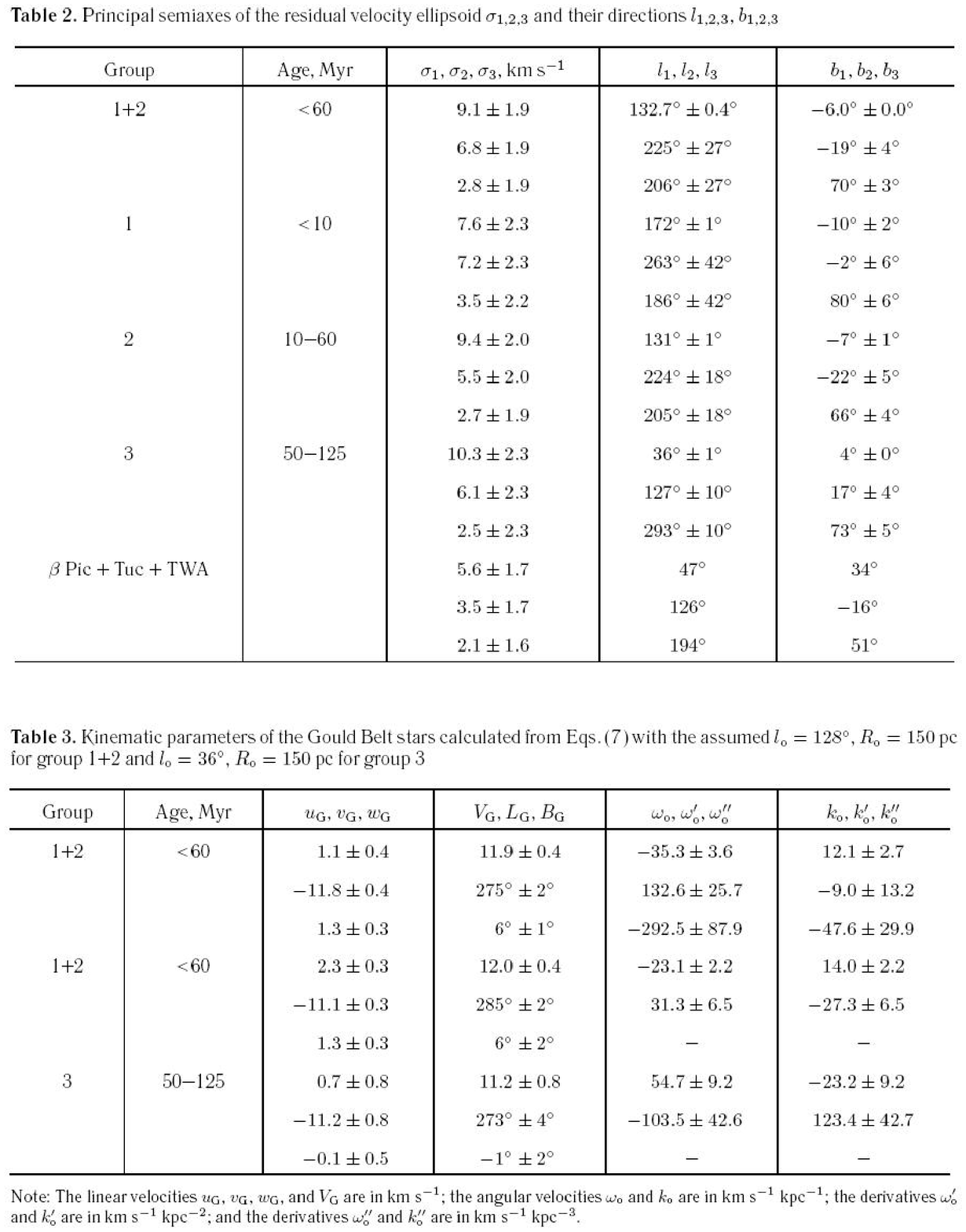}
\end{center}}
\end{figure}

\begin{figure}[t]
{\begin{center}
  \includegraphics[width=78mm]{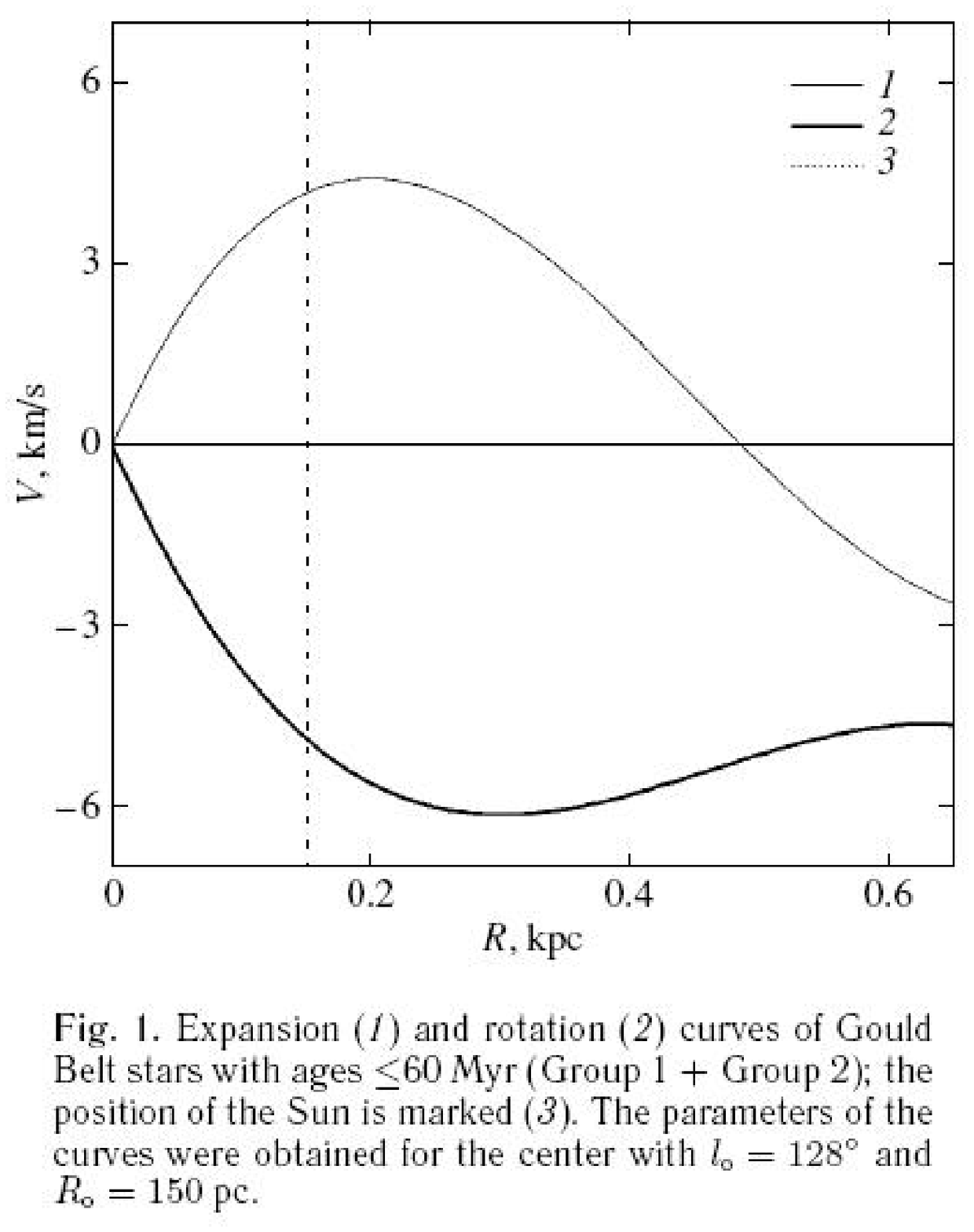}

  \includegraphics[width=160mm]{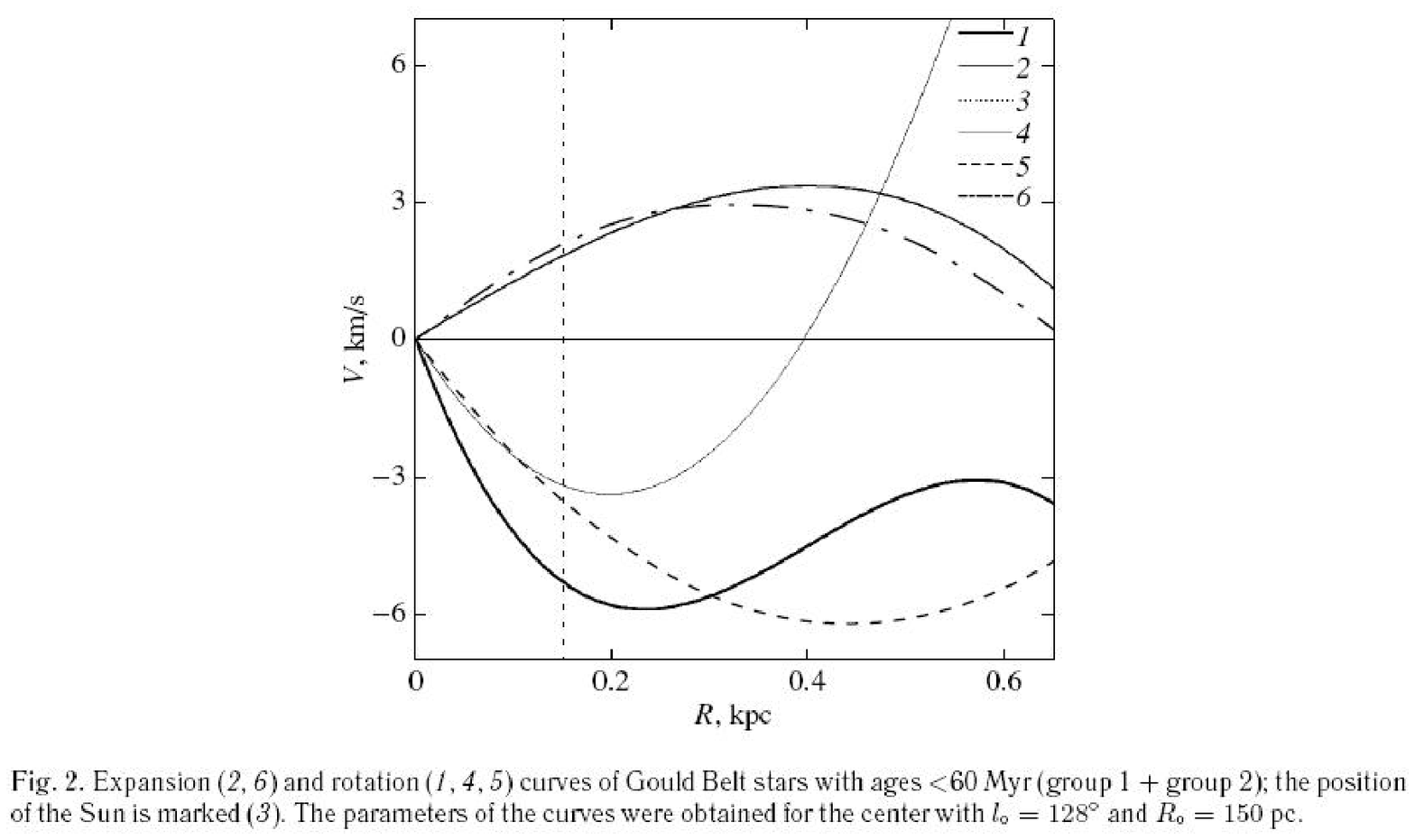}
\end{center}}
\end{figure}

\begin{figure}[t]
{\begin{center}
  \includegraphics[width=160mm]{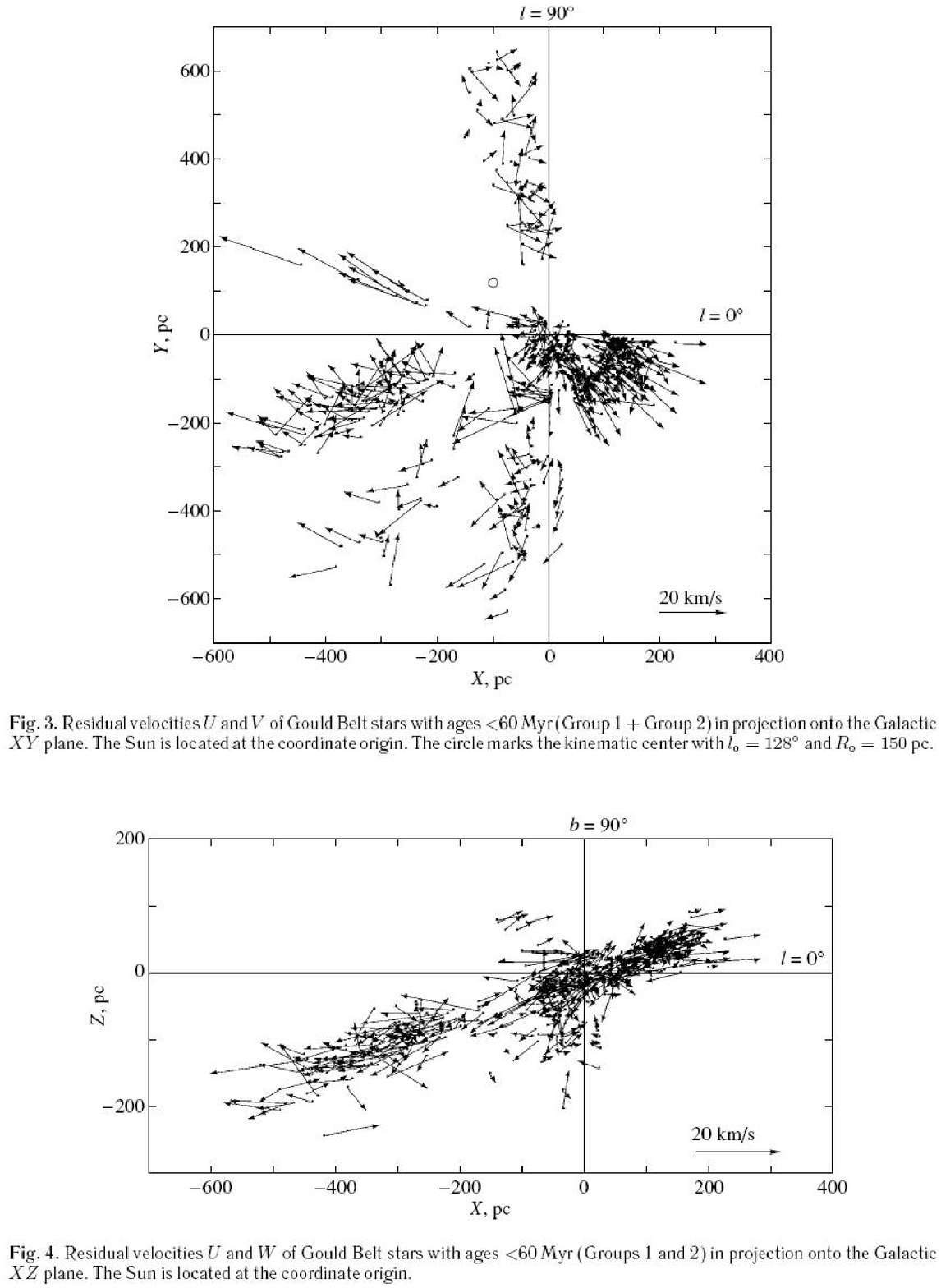}
\end{center}}
\end{figure}

\begin{figure}[t]
{\begin{center}
  \includegraphics[width=78mm]{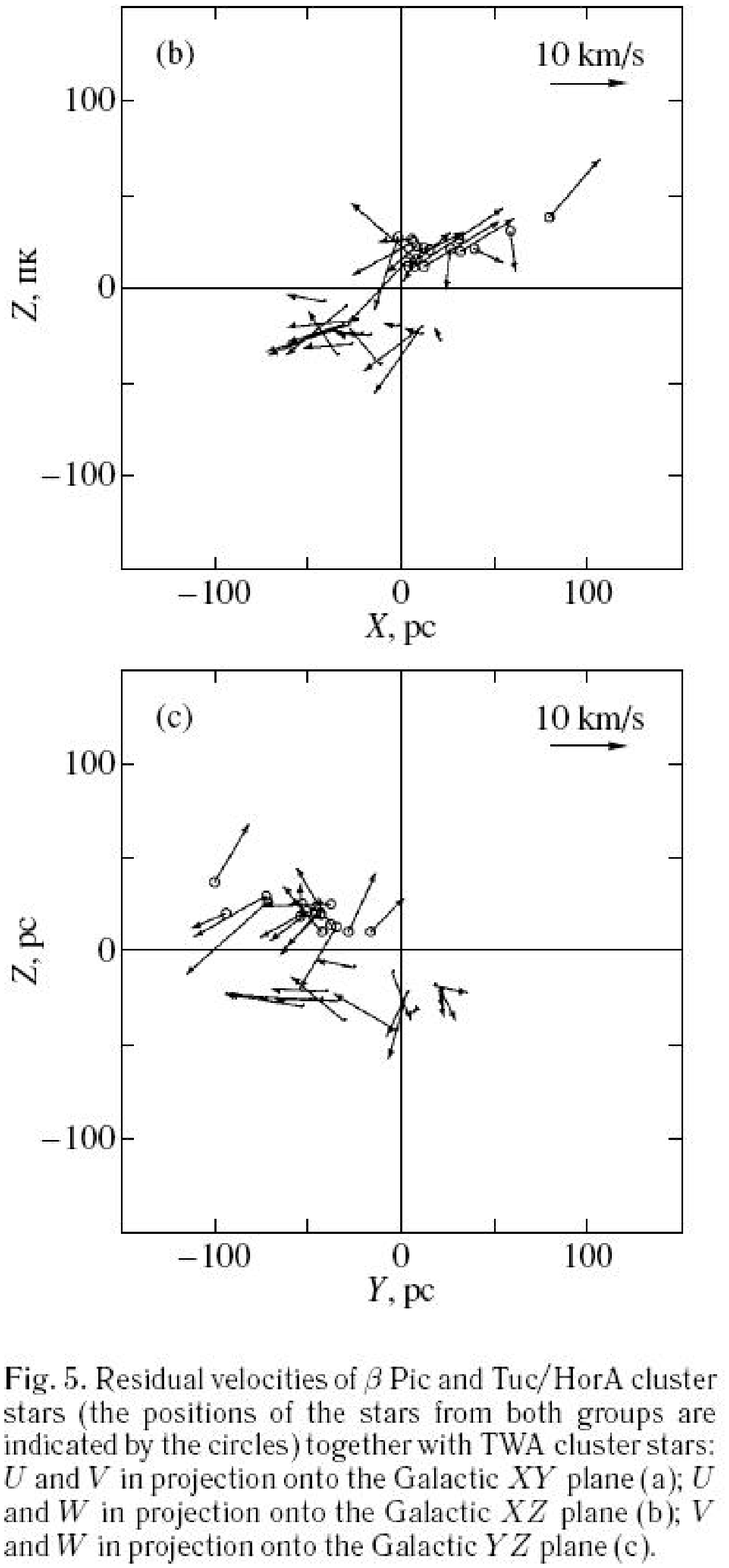}
\end{center}}
\end{figure}

\end{document}